\documentclass[11pt]{article}
\usepackage{moriond,epsfig}

\def\be{\begin{equation}}
\def\ee{\end{equation}}
\def\bea{\begin{eqnarray}}
\def\eea{\end{eqnarray}}

\begin{document}
\vspace*{4cm}
\title{SEARCHES FOR NEW PHYSICS IN $ep$ SCATTERING AT HERA}
\author{D.~M.~SOUTH}
\address{{\it (on behalf of the H1 and ZEUS collaborations)} \\
Deutsches Elektronen Synchrotron\\
Notkestrasse 85, 22607, Hamburg, Germany\\
E-mail: David.South@desy.de }

\maketitle\abstracts{The latest results from the H1 and ZEUS collaborations
are presented on leptoquark production and rare Standard Model processes.
The data were taken in the period 1994--2005, at a centre of
mass energy of up to 319 GeV. Intriguing events containing isolated leptons
and missing transverse momentum, as well as multi--lepton events, are observed by H1
in regions of phase space where the SM prediction is low. Interpretations of the 
observed excesses in terms of physics Beyond the Standard Model are also discussed.}

\section{Introduction}
A comprehensive physics programme is employed by the H1 and ZEUS
experiments at the HERA $ep$ collider. To date, each experiment has collected
over 300 pb$^{-1}$ of data for physics analysis, consisting of approximately
equal amounts of $e^{+}p$ and $e^{-}p$ scattering. Together with measuring the structure
of the proton, the deep inelastic collisions (DIS) produced at HERA at a centre
of mass up to 319~GeV provide an ideal environment to study rare processes.

Model dependent searches are employed to set constraints on the Standard Model (SM)
and look for specific new particles and physics beyond the Standard Model (BSM). The
search for the production of leptoquarks at HERA, including lepton flavour violating
models, is described in section \ref{sec:lq}. Model independent searches are employed
at HERA to look for anomalies in the high transverse momentum ($P_{T}$) phase space sparsely
populated by the SM, the results of which are presented in sections
\ref{sec:general}--\ref{sec:multilep}. A summary is given in section \ref{sec:summary}.

\section{Leptoquark Production and Lepton Flavour Violation}
\label{sec:lq}

The $ep$ collsions at HERA provide a unique possibility to investigate
the formation of a new particle coupling to a lepton--quark pair. In the
Buchm\"{u}ller, R\"{u}ckl and Wyler (BRW) classification of such
states, termed ``leptoquarks'', 7 scalar and 7 vector particles are proposed,
all of which can couple to an $eq$ pair and 4 of which can also couple to both
$eq$ and $\nu q$~\cite{brw}. Leptoquarks are hence bosons of fractional charge
that carry both leptonic ($L$) and baryonic ($B$) numbers, such that their fermion
number ($F=3B+L$) can be $F=0$ or $F=2$. A leptoquark that couples to all lepton
flavours would result in the possibility of lepton flavour violating (LFV) processes
in $ep$ collisions.

\begin{figure}[h]
  \centering
  \includegraphics[width=.45\textwidth]{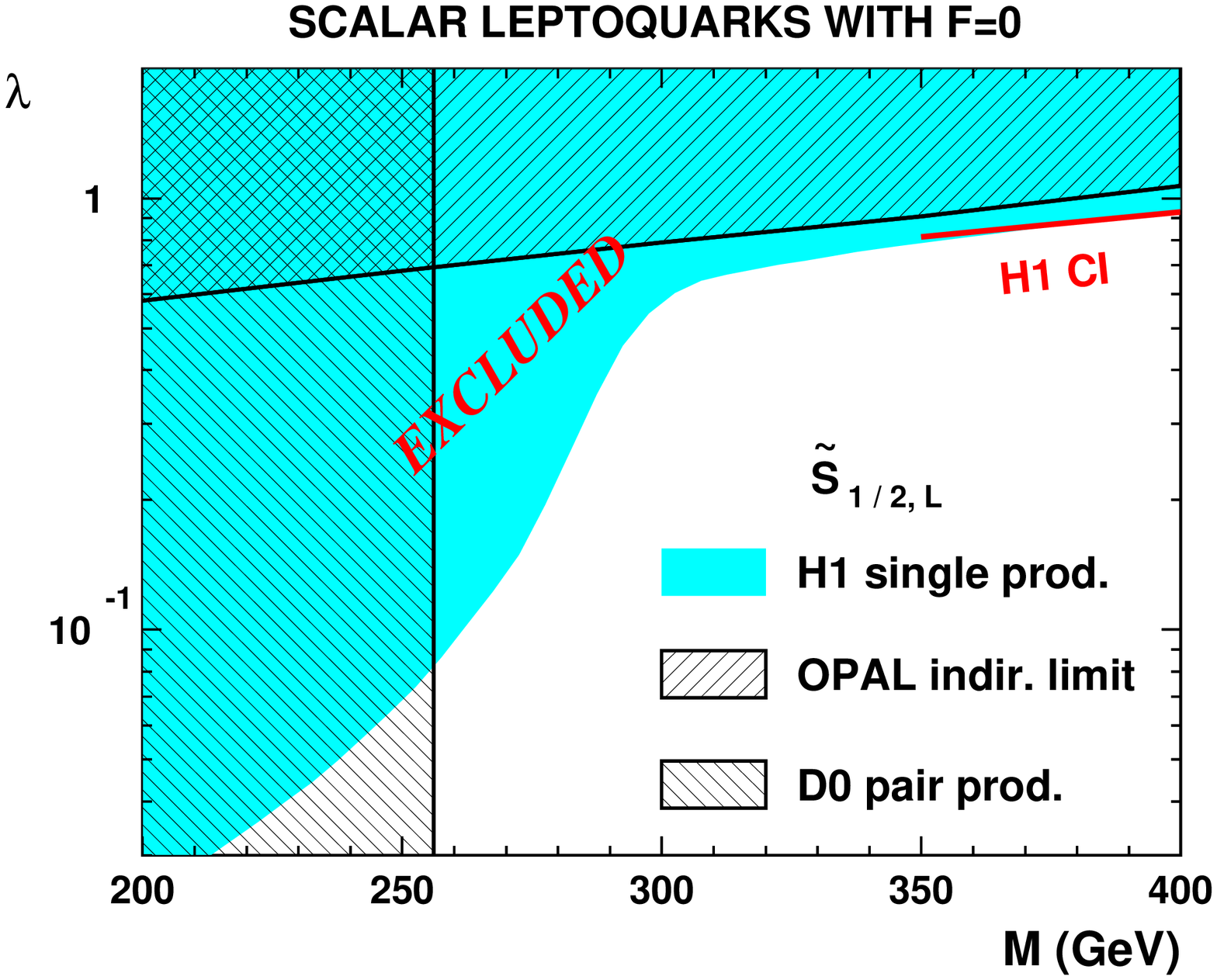}
  \includegraphics[width=.53\textwidth]{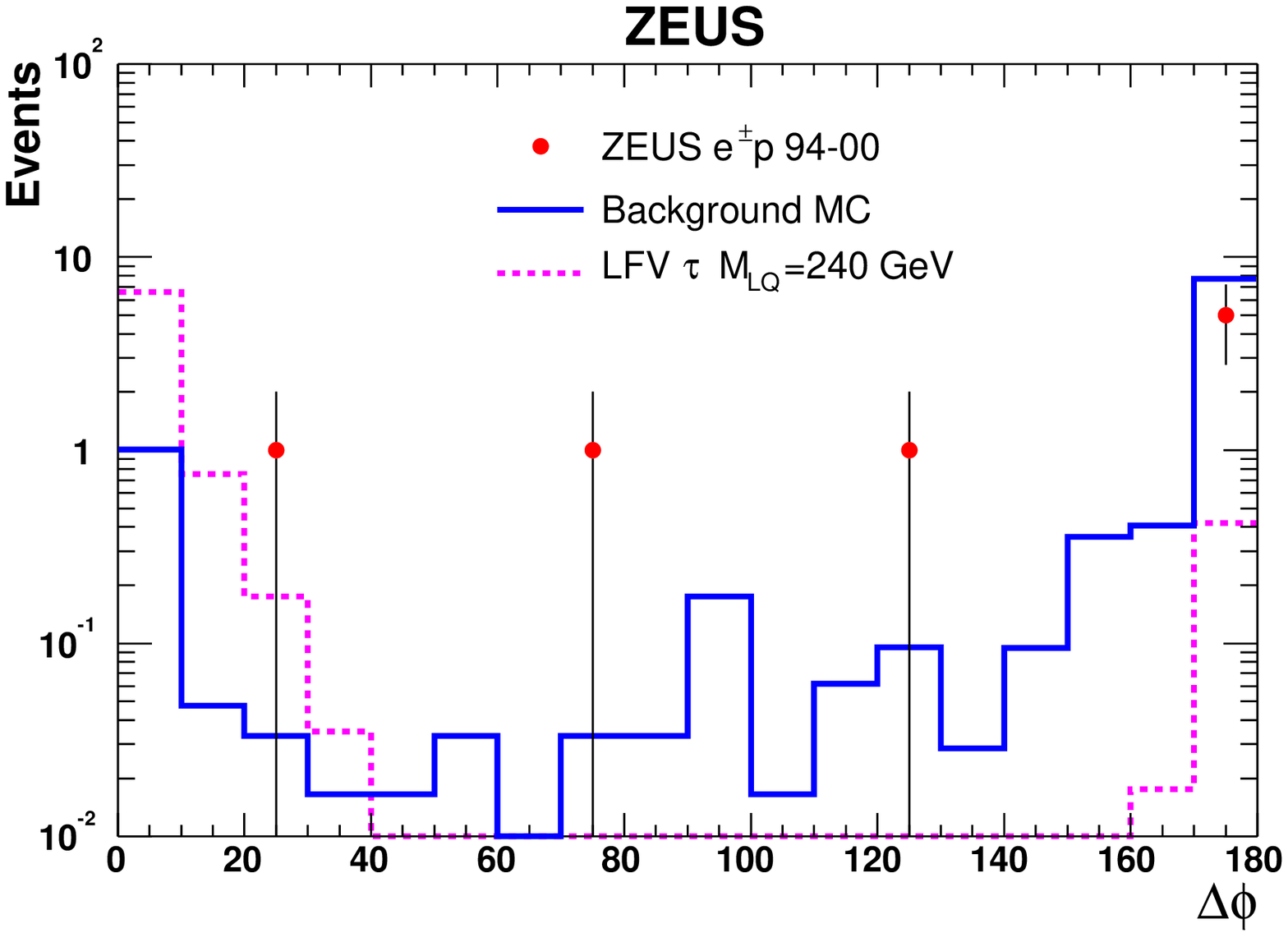}
  \caption{Left: H1 exclusion limits at 95\% CL on the coupling $\lambda$ as a function of
	leptoquark mass for $F=0$ scalar leptoquarks in the framework of the BRW model.
	The derived limits are compared to those from the D0 and OPAL experiments. Right:
	The difference in azimuthal angle of the candidiate lepton and missing $P_{T}$, as observed
	at the preselection level of the ZEUS LFV leptoquark search. The data are the points,
	the solid histogram is the SM prediciton and the dashed line represents the signal
	expectation with arbitrary normalisation.}
\label{fig:leptoquarks}
\end{figure}

A search for first generation leptoquark production, $eq\rightarrow eq$
or $eq\rightarrow \nu q$ has been performed on the HERA~I data by the H1~\cite{h1leptoquark}
and ZEUS~\cite{zeusleptoquark} collaborations by looking for deviations in the mass
spectra of neutral current (NC) and charged current (CC) DIS interactions, the major SM
background. No evidence of leptoquark production is observed, and limits on the
14 leptoquark couplings are derived as function of mass, as shown for example from H1
for $F=0$ scalar leptoquarks in figure \ref{fig:leptoquarks} (left), where the derived
limits are compared to those from the D0~\cite{d0lq} and OPAL~\cite{opallq} experiments. 
Leptoquark masses up to 386~GeV are ruled out at 95\% confidence level (CL) for a coupling
$\lambda$ of electromagnetic strength ($\lambda$~=~$\sqrt{4 \pi \alpha_{em}} =$~0.3).

A search for LFV processes mediated by leptoquark exchange
has also been performed by H1~\cite{h1leptoquarklfv} and ZEUS~\cite{zeusleptoquarklfv},
investigating the interactions $eq\rightarrow \mu q$ and $eq \rightarrow \tau q$. The
analysis has a low SM background and thus high sensitivity to such processes. Figure
\ref{fig:leptoquarks} (right) shows the $\Delta\phi$ distribution at the preselection level
of the ZEUS analysis, where $\Delta\phi$ is the difference in azimuthal angle of the lepton and
missing $P_{T}$ in the event. A cut is subsequently made in the final selection of
$\Delta\phi <$~20$^{o}$. It can be seen that good separation of the LFV leptoquark production signal
and the SM background is achieved, although no evidence for LFV is observed. Similarly to the
above analyses, limits are derived for the appropriate leptoquark couplings as a function of mass.
The ZEUS analysis of the $eq \rightarrow \tau q$ channel, which uses 113~pb$^{-1}$ of $e^{+}p$ data,
improves the existing limits from rare $\tau$, $B$ or $K$ decays (see references
given in~\cite{zeusleptoquarklfv}).

\section{A General Search for New Phenomena}
\label{sec:general}

A model independent general search for deviations from the SM has been performed by
H1~\cite{general}, using a HERA~I data sample corresponding to an integrated luminosity of
117~pb$^{-1}$. All high $P_{T}$ final state configurations involving electrons ($e$),
muons ($\mu$), jets ($j$), photons ($\gamma$) or neutrinos ($\nu$) are considered.
All final state configurations containing at least two such objects with $P_{T} >$~20~GeV
in the central region of the detector are investigated and classified into exclusive
event classes, $e$--$j$, $\mu$--$j$--$\nu$, $j$--$j$--$j$ and so on.

\begin{figure}[t]
  \centering
  \includegraphics[height=.44\textheight]{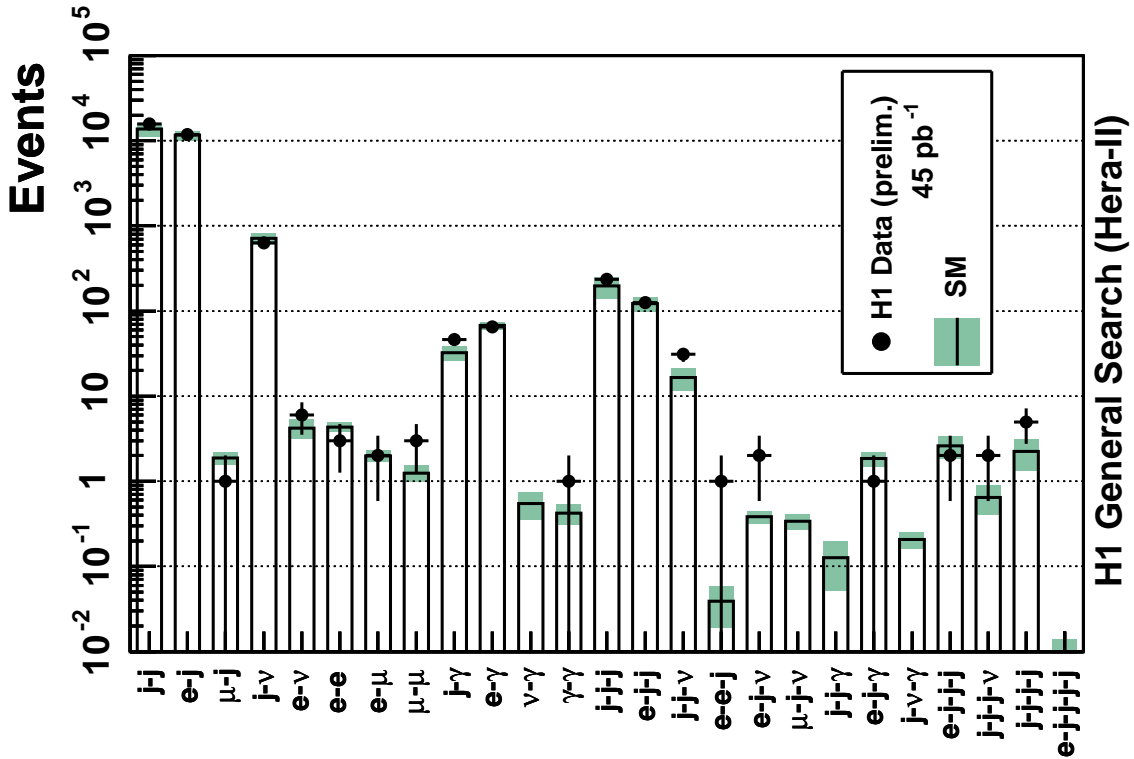}
  \caption{Event yields for all event classes with a SM expectation greater than 0.01 events in the
	H1 HERA~II general search analysis. The data are the points, the histogram bars are the SM
	prediction and the shaded band is the total SM error.}
\label{fig:general}
\end{figure}

Data events are found in 22 such event classes and a good agreement is observed between
data and the SM expectation in most event classes. A non--biased statistical method is employed
to search for deviations of the data with respect to the SM. A good agreement is found in all channels,
except in the $\mu$--$j$--$\nu$ event class, where 4 data events are observed compared
to a SM expectation of 0.8~$\pm$~0.2 as previously reported in~\cite{isoleph1newwpaper}.
Additionally, in the $e$--$j$--$j$--$j$--$j$ event class 1 event is observed in the data compared
to a SM prediction of 0.026~$\pm$~0.011.

The H1 general search has been repeated using 45 pb$^{-1}$ of HERA~II $e^{+}p$ data~\cite{multilep},
the results of which are shown in figure \ref{fig:general}. As in the HERA~I analysis, a good overall
agreement is observed between the data and the SM prediction and data events are observed in 20 of the
event classes. The deviation of the data with respect to the SM  in the $e$--$j$--$\nu$ event class
corresponds to the observed excess in this data sample as reported in~\cite{isoleph1hera2}. The observed
excesses in the $e$--$j$--$\nu$ and $\mu$--$j$--$\nu$ event classes, in which the main SM contribution
arises from real $W$ production, are described in more detail in section \ref{sec:isolep}.

\section{Events containing Isolated Leptons and Missing Transverse Momentum}
\label{sec:isolep}

Events containing a high $P_{T}$ isolated electron or muon and large missing transverse
momentum have been observed at HERA~\cite{isoleph1newwpaper,isoleph1origwpaper,isolepzeusorigwpaper}.
The main SM contribution to such a topology comes from the production of real $W$ bosons
$ep \rightarrow eW^{\pm}X$ with subsequent leptonic decay $W \rightarrow l\nu$. An excess
of HERA~I (1994--2000) data events compared to the SM prediction was reported by the H1
collaboration~\cite{isoleph1newwpaper}, which was not confirmed by the ZEUS collaboration,
although using a slightly different analysis approach~\cite{zeustop}.

\begin{figure}[h]
  \centering
  \includegraphics[width=.49\textwidth]{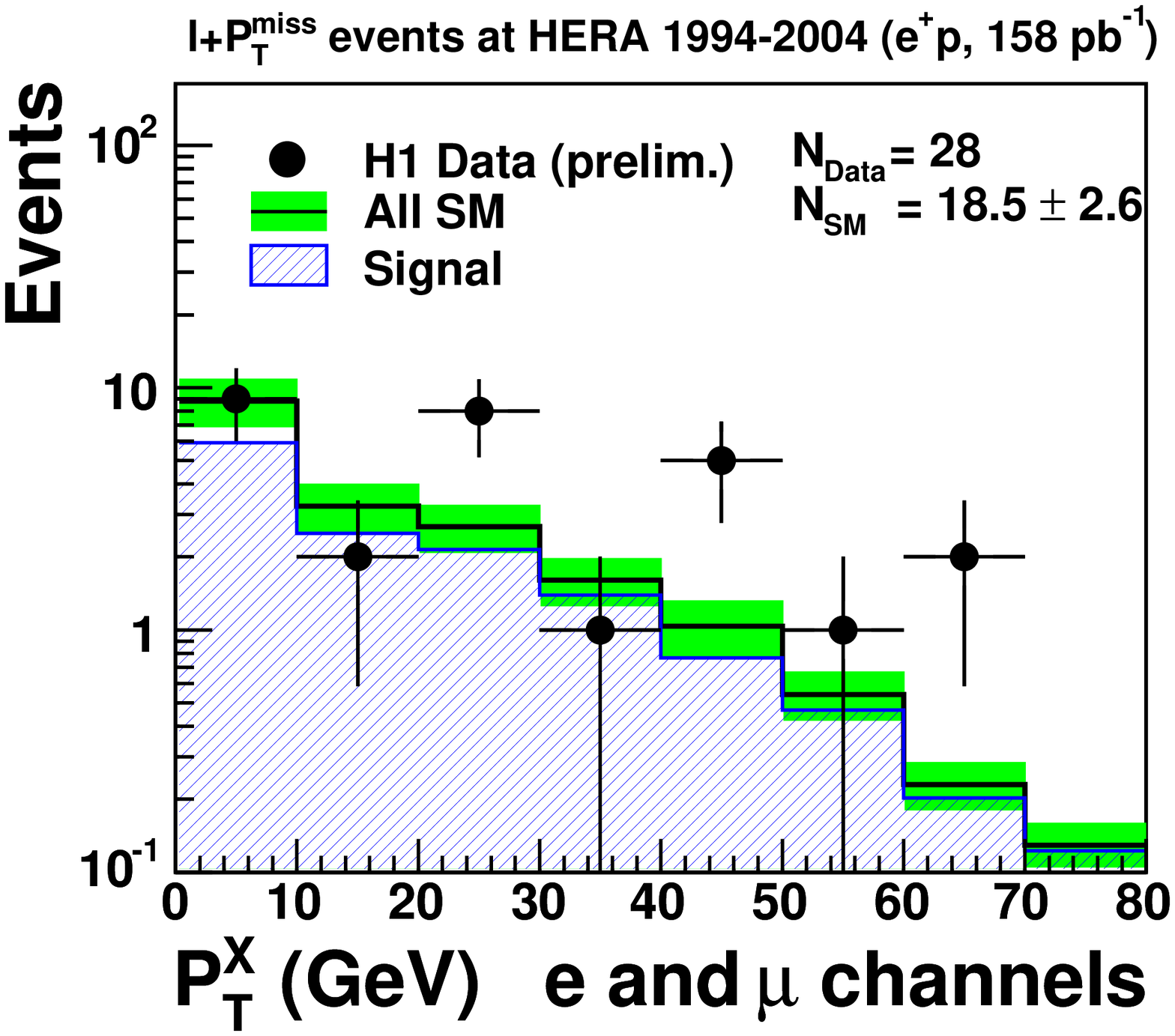}
  \includegraphics[width=.49\textwidth]{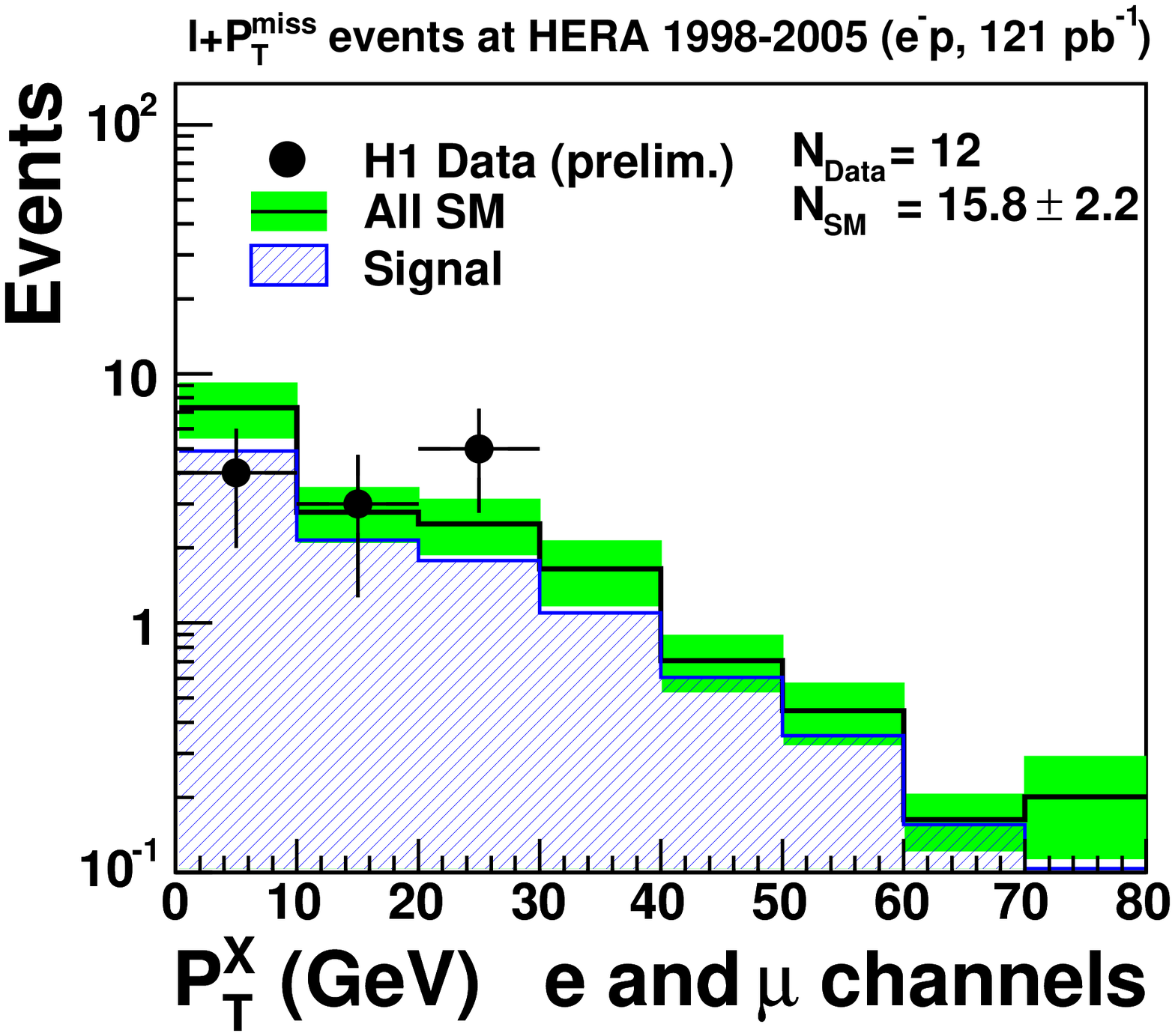}
  \caption{The hadronic transverse momentum spectra of the observed events in the H1
	isolated lepton analysis. The sample is divided into the $e^{+}p$ (shown on
	the left, ${\cal L} =$~158~pb$^{-1}$) and $e^{-}p$ (right, ${\cal L} =$~121~pb$^{-1}$)
	data samples. The data are the points, the full histogram is the SM expectation and
	the shaded band is the total SM error. The signal component, dominated by real $W$
	production, is shown by the hatched histogram.}
\label{fig:isolep}
\end{figure}

The H1 analysis has been updated~\cite{isoleph1hera2,isoleph1hera2new} to include new
$e^{\pm}p$ data from the ongoing HERA~II phase (2003--2005), resulting in a total analysed
luminosity of 279~pb$^{-1}$. A total of 40 events are observed in the data, compared to a SM
prediction of 34.3~$\pm$~4.8. The hadronic transverse momentum ($P_{T}^{X}$) spectra of the 
$e^{\pm}p$ data are presented in figure \ref{fig:isolep}. At large values of $P_{T}^{X}$
an excess of $e^{+}p$ data events is observed compared to the SM expectation, as can be seen
in figure \ref{fig:isolep} (left). For $P_{T}^{X} >$~25~GeV a total of 15 data events are
observed compared to a SM prediction of 4.6~$\pm$~0.8. The observed excess is equivalent to
a fluctuation of approximately 3.4$\sigma$. Interestingly, the excess is not observed in the
current $e^{-}p$ analysis, as can be seen in figure \ref{fig:isolep} (right), which includes
an almost factor of 10 increase in statistics with respect to the HERA~I $e^{-}p$ data set.
The results of the H1 analysis are summarised in table 1.%%\ref{tab:isolep}. DOESN'T WORK!!?

A re-analysis of the ZEUS electron channel has been performed~\cite{isolepzeushera2},
using $e^{+}p$ data from 1999--2000 in addition to HERA~II $e^{+}p$ data from 2003--04
(total luminosity 106~pb$^{-1}$), leading to a similar SM background expectation to the H1
analysis. However, the ZEUS analysis does not confirm the observed H1 excess, with only one
electron candidate observed in the region $P_{T}^{X} >$~25~GeV compared to a SM prediction of
1.5 ~$\pm$~0.2. It should be noted, however, that the two analyses remain different in
terms of phase space; in particular the ZEUS analysis has a more restrictive
polar angle range. 

\begin{table}[t]
\label{tab:isolep}
\caption{Summary of the H1 results of searches for events with isolated electrons or muons and missing
	transverse momentum for the $e^{+}p$ data ($\cal L$=158 pb$^{-1}$), $e^{-}p$ data
	($\cal L$=121 pb$^{-1}$) and the full HERA data set ($\cal L$=279 pb$^{-1}$), in the
	region $P_{T}^{X}>25$~GeV. The number of observed events are compared to the SM prediction.}
\vspace{0.4cm}
\begin{center}
\begin{tabular}{|c||c|c|c|}
\hline
H1 Preliminary & e channel & $\mu$ channel & combined e \& $\mu$\\
$P_{T}^{X} >$ 25 GeV & & & \\
\hline
\hline
1994--2004 $e^{+}p$   158 pb$^{-1}$ &  9 / 2.3 $\pm$ 0.4 & 6 / 2.3 $\pm$ 0.4 & 15 / 4.6 $\pm$ 0.8 \\
\hline
1998--2005 $e^{-}p$   121 pb$^{-1}$ &  2 / 2.4 $\pm$ 0.5 & 0 / 2.0 $\pm$ 0.3 &  2 / 4.4 $\pm$ 0.7 \\
\hline
\hline
1994--2005 $e^{\pm}p$ 279 pb$^{-1}$ & 11 / 4.7 $\pm$ 0.9 & 6 / 4.3 $\pm$ 0.7 & 17 / 9.0 $\pm$ 1.5 \\
\hline
\end{tabular}
\end{center}
\end{table}

\begin{figure}[h]
  \centering
  \includegraphics[width=.45\textwidth]{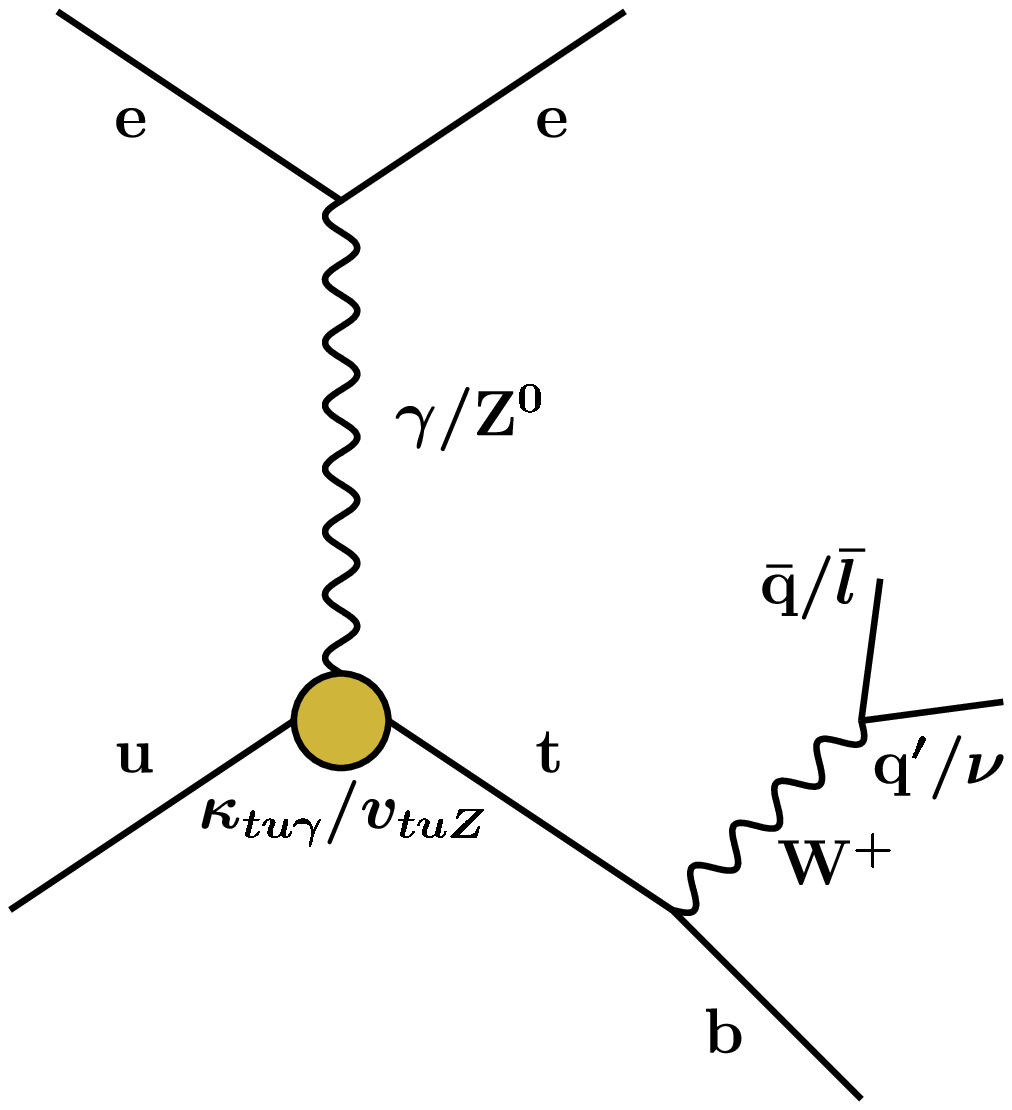}
  \includegraphics[width=.53\textwidth]{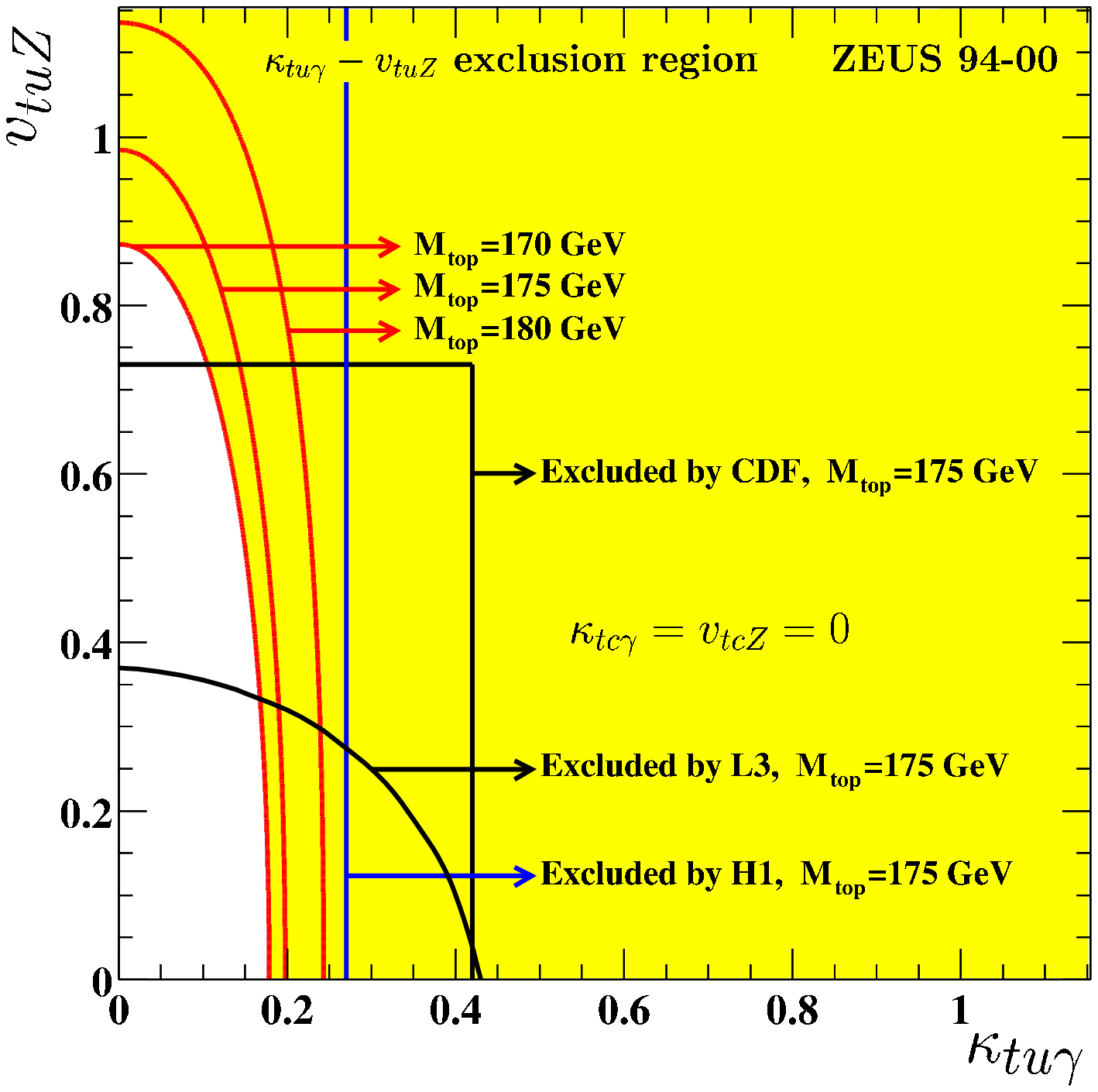}
  \caption{Left: Single top production at HERA via FCNC. Right: Exclusion limits at 95\% CL on the
	anomalous $\kappa_{tu\gamma}$ and $v_{tuZ}$ couplings derived by H1 and ZEUS, compared to limits
	derived by the CDF and the L3 experiments.}
\label{fig:singletop}
\end{figure}

Both experiments have also searched for the non--standard production of single top quarks at
HERA~I via the Flavour Changing Neutral Current (FCNC) process, with subsequent leptonic $W$
decay~\cite{zeustop,h1top}. The analysis strategy is an extension of the isolated lepton analysis
described above, where the topology, as illustrated in figure \ref{fig:singletop} (left), could
explain the observed H1 excess (although this process would predict a similar signal rate in $e^{+}p$
and $e^{-}p$ collisions). While some of the H1 events at large $P_{T}^{X}$ show kinematics
compatible with the single top hypothosis, no signal can be claimed. Figure \ref{fig:singletop}
(right) shows limits derived at 95\% CL by both experiments on the anomalous FCNC $\kappa_{tu\gamma}$
coupling, compared to limits derived by the CDF~\cite{cdftop} and L3~\cite{leptop} experiments.
HERA has the best sensitivity to the $\kappa_{tu\gamma}$ coupling in the region where $v_{tuZ}$ is small.
The ZEUS limits are also derived with a dependence on the $v_{tuZ}$ coupling.

\section{Multi--lepton Events }
\label{sec:multilep}

Searches for multi--electron production at high transverse momentum have been previously
carried out by the H1~\cite{multie} and ZEUS~\cite{multiezeus} experiments, using the
HERA~I data sample. The production of high $P_{T}$ muon
pairs has also been studied at HERA~I by both H1~\cite{multimu} and ZEUS~\cite{multimuzeus}.
The main SM process for multi--lepton production in $ep$ collisions is
photon--photon interactions $\gamma\gamma\rightarrow l^{+}l^{-}$, where quasi real
photons radiated from the incoming electron and proton interact to produce a pair of
leptons. High mass events ($M_{1,2} >$~100~GeV) are observed in the in the di--electron
sample of both experiments, and additionally in the H1 tri--electron sample, regions where
the SM expectation is low.

\begin{figure}[h]
  \centering
  \includegraphics[width=.87\textwidth]{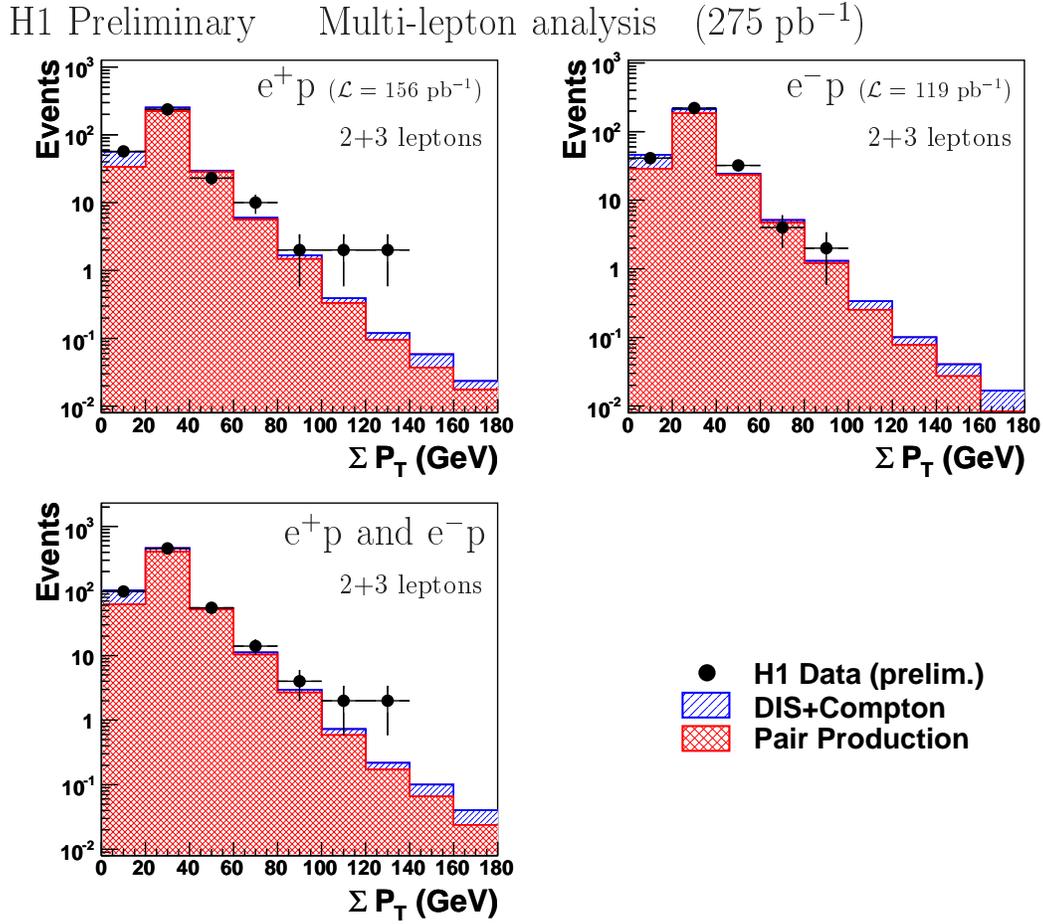}
  \caption{Distributions of the scalar sum of the transverse momenta of the combination of
	all di-- and tri--lepton events for data taken in $e^{+}p$ (upper left) and $e^{-}p$ (upper right) collisions.
	The combination of all di-- and tri--lepton events for all HERA data is shown in the lower figure.
	The data (points) are compared to the SM expectation (histogram), which is dominated by pair production.	
	}
\label{fig:multilep}
\end{figure}

H1 has recently updated the analysis to include the new HERA~II data~\cite{multilep,multilepnew},
now exploiting a total luminosity of 275~pb$^{-1}$. The analysis examines $e\mu$, $\mu\mu$,
$e\mu$, $eee$ and $e\mu\mu$ topologies, searching for events with at least two high
$P_{T}$ electrons or muons. Figure \ref{fig:multilep} shows the scalar sum of the
transverse momenta of the selected events containing two or three high $P_{T}$ leptons,
for the positron and electron data separately, and for all HERA together. The data are found
to be in good overall agreement with the SM, although interesting events are seen at high masses and,
similarly to the analysis described in section \ref{sec:isolep}, only in the $e^{+}p$ data.
The high mass events include di--electron and tri--electron events observed in the HERA~I data
and, more recently, electron--muon events observed in the HERA~II data.

The production in $ep$ collisions of a doubly charged Higgs boson $H^{++(--)}$ could be
a source of events containing multiple high $P_{T}$ leptons and the observed high mass
events in the H1 analysis have been investigated in this context~\cite{doublehiggs}.
Only one $ee$ event satisfies the additional selection criteria and the
HERA limits on the $H^{++(--)}$ coupling to $ee$ are not competitive to those set by
the OPAL experiment~\cite{opalhiggs}. However, limits derived from the HERA data in
the $e\mu$ decay channel of the $H^{++(--)}$ extend to higher masses beyond the reach
of the previous searches performed by the CDF~\cite{cdfhiggs} and LEP~\cite{lephiggs}
experiments and new constraints are obtained for the $H^{++(--)}$ coupling to $e\tau$.

\section{Summary}
\label{sec:summary}

Many searches for new physics have been performed at HERA by the H1 and ZEUS
collaborations. No evidence for the production of leptoquarks is observed.
Interesting events containing isolated leptons and missing $P_{T}$ as well as multiple
high $P_{T}$ leptons at high masses are observed by H1. The continued data taking
in the HERA~II phase by both experiments will hopefully clarify the high $P_{T}$ lepton
events seen by H1 and further the search for new physics at HERA.

\section*{Acknowledgments}
I would like to thank Martin Wessels, who deserves full credit for composing the
presented talk and who's misfortune allowed me to attend such a wonderful conference.
Thanks also to Guillaume Unal for providing much needed help in arranging my arrival
at La Thuile.

\end{document}